\documentclass[preprint,10pt,a4paper]{aastex}
\usepackage{amsmath}
\usepackage{graphicx}
\usepackage{multicol}
\newcommand{\be}{\begin{equation}} 
\newcommand{\ee}{\end{equation}}
\newcommand{\nn}{\mbox{} \nonumber \\ \mbox{} }
\newcommand{\ba}{\begin{eqnarray}}
\newcommand{\ea}{\end{eqnarray}}

\newcommand{\E}{{\bf E}}
\newcommand{\B}{{\bf B}}

\newcommand\eg{\textit{e.g.,\ }}

\newcommand{\Bf}{{magnetic field}}
\newcommand{\Bfs}{{magnetic fields}}
\newcommand{\Ef}{{electric  field}}
\newcommand{\Efs}{{electric fields}}

\newcommand{\NS}{neutron star}
\newcommand{\NSs}{{neutron stars}}
\newcommand{\EM}{electromagnetic}
\newcommand{\BH}{{black hole}}
\newcommand{\BHs}{{black holes}}
\newcommand{\Sc}{Schwarzschild}
\newcommand{\ms}{magnetosphere}
\newcommand{\mss}{magnetospheres}

\begin{document}

\title{The electromagnetic model of short GRBs,  the nature of prompt tails,  supernova-less long GRBs   and  highly efficient episodic accretion}

\author{Maxim Lyutikov\\
Department of Physics, Purdue University, \\
 525 Northwestern Avenue,
West Lafayette, IN
47907-2036 }

\begin{abstract}
Many short GRBs show   prompt tails  lasting up to hundreds of seconds that  can  be energetically dominant over the initial  sub-second spike. 
In this paper we develop an electromagnetic model of short GRBs that explains  the two stages of the  energy release, the prompt spike and the prompt tail. 
The key ingredient of the model  is the recent discovery that  an {\em  isolated}   black hole   can keep its open magnetic flux  for times much longer than the collapse time and, thus, can spin-down electromagnetically,  driving the  relativistic wind. 

First, the merger is preceded by an \EM\  precursor  wind with total   power $L_p \approx { ( G M_{NS})^3 B_{NS}^2 \over c^5 R} \propto (-t)^{-1/4}$,  reaching  $ 3 \times 10^{44}  \, {\rm erg\,  s}^{-1} $ for typical \NS\ masses of $1.4 M_\odot$ and \Bfs\ $B\sim 10^{12} $ G. If a fraction of this power is converted into pulsar-like coherent radio emission, this may produce observable radio burst of few milliseconds (like the Lorimer burst).

At the active stage of the merger,  two neutron stars produces  a black hole  surrounded by an accretion torus in which the 
magnetic field is amplified to $\sim   10^{15}$ Gauss. This magnetic field  extracts the rotational energy of the black hole and drives an axially-collimated electromagnetic wind that  may carry of the order of  $10^{50}$ ergs, limited by the accretion time of the torus, a few hundred milliseconds. For observers nearly aligned with the orbital normal this is seen as a classical short GRB. 

After the  accretion of the torus,  the isolated  black hole  keeps the open magnetic flux and  drives the equatorially (not axially) collimated outflow, which is seen by an  observer at intermediate polar angles  as a prompt tail. The tail carries more energy than the prompt spike, but its emission is de-boosted for observers along  the orbital normal. Observers in the equatorial plane miss the prompt spike and interpret the prompt tail as an energetic  long GRB (the  supernova-less long burst GRB060505 and GRB060614 may belong to this category).

We also demonstrate that episodic accretion onto the \BH\ of magnetized clouds that carry non-zero magnetic flux can be highly efficient in extracting the spin energy of the \BH,  producing the \EM\ outflows with the power  exceeding  the average $\dot{M} c^2$  accretion power and total  energy exceeding the rest mass energy of the accreted mass. We identify the late time flares with such accretion  events.

%The model argues that the early $X$-ray GRB afterglows, at times $t \leq 10^{5}$ seconds come mostly from internal dissipation in the wind, and not from the forward shock (which does dominate at later times in optical and radio wavebands). 
\end{abstract}

\section{Challenges of short GRBs}

We are confident that  GRBs come from the release of gravitational energy (or rotational energy, which is smaller)  during  collapse of a stellar mass object \citep[for a recent review, see][]{GehrelReview}.
As a first approximation we can  then associated the time scale of the dominant energy release with an average density, 
$t \geq 1/\sqrt{G \rho}$ in the pre-GRB object. For  for short GRBs this implies $\rho \geq  10^7$ g cm$^{-3}$;  leaving  the neutron stars as the main candidate for short GRBs. (Somewhat longer times  are related to the accretion time scale of a torus  formed during  \NS-\NS\ mergers, see below).

The association of short GRBs with the \NS-\NS\ mergers is far from proven \citep[see, \eg  a critique of modern GRB models by][]{Lyutikov:2009}. The most contradictory observation, in our view, is that   a  number of short GRBs show an extended emission tail \cite[\eg][]{2010ApJ...717..411N} and flares. This is one of the most important problem in the NS-NS merger paradigm for  short GRBs: the release of a substantial and often dominant amount of energy on times scales millions of  times longer than the dynamical time scale.
  Resolution of this problem  might lead to the identification of the short GRB  progenitors. We offer  a possible resolution in the present paper. 

  Numerical simulations indicate that the active stage of NS-NS coalescence typically takes  a very short time, 10 msec  \citep[\eg][]{Kiuchi}. 
  Highly spinning \NSs\ with hard equation of state may  extend the collapse time to 100 msec \citep{2011PhRvD..83l4008H,2010CQGra..27k4105R,2011PhRvL.107e1102S}. 
  Under certain conditions (the considerable mass asymmetry and a specific range of the total mass of the system) the merger of two \NSs\ can lead to the formation of a massive, $\leq 0.35 M_\odot$, disk \citep{2010CQGra..27k4105R}. The disk typically accretes on time scale of tens to hundreds of milliseconds \cite[\eg][]{2011ApJ...732L...6R}, so that after few seconds the mass accretion rate is orders of magnitude smaller \citep{2008MNRAS.390..781M}. 
   A small amount   of material, typically $\leq 10^{-2}- 10^{-3} M_\odot$,  may also be ejected during the merger of   unequal-mass \NSs, producing tidal tails that  accrete on time-scales of 1-10 secs, depending on the assumed $\alpha$ parameter of the disk \cite[\eg][]{Kiuchi, LiuNSNS,2007MNRAS.376L..48R,Faber,2011ApJ...734...35C}. 
  
To summarize the relevant results of the numerical simulations, the active part of the coalescence may last up to hundreds of milliseconds, but not much longer. 
 Thus, any energetically dominant  activity on much longer time scales  formally  contradicts the NS-NS coalescence paradigm  for short GRBs.  Most puzzling are 
 cases when the tail fluence  dominates over the primary burst \citep[by a factor of 30 as in GRB080503][]{Perley}, or when powerful flares appear late in the afterglow \citep[\eg   in case of GRB050724 there is a powerful flare at $10^5$ sec][]{2005ApJ...635L.133B,2006A&A...454..113C}. In the standard forward shock model of afterglows this requires that at the end of the activity, lasting 10-100 msec,  the source releases  more energy than during the prompt emission  in a form of a low $\Gamma$ shell, which collides with the FS  after $\sim 10^6$ dynamical times, a highly fine-tuned  scenario.

The latest Fermi results \citep[\eg\ on GRB080916C,][]{FermiGRB080916C} show a  similarity of the  long lasting very high energy  emission of  the short and long  GRBs. GeV signals from both types of bursts start with a short delay, of fewer than several seconds,  with respect to prompt emission and continues for hundreds to thousands seconds after the prompt emission.  Such continuity, from nearly simultaneous with prompt to long lasting emission in both cases is quite  surprising, especially  if photons are produced in the forward shock with drastically different properties for long and short GRBs.

Previously, the extended emission in shorts GRBs was attributed either to long lived magnetar-type central object  \citep{2012MNRAS.419.1537B} or to a build up of the sufficiently large magnetic flux  on the central \BH\ \citep{2011MNRAS.417.2161B}, which becomes nearly independent of the mass accretion rate and can ensure high extracted power even though  the mass accretion rate falls off sharply, on a time scale of hundreds of milliseconds. In case of the magnetar-powered extended emission, it is not clear if a merger of two \NSs\ can create a sufficiently long-lived (on a time scale of hundreds of seconds) object and whether the wind can be made sufficiently clear of the baryons, while in the accretion model it s not clear if the disk provides sufficient accretion power on such long time scales.

In this paper we develop a model of the extended emission in short GRBs that is reminiscent of the magnetar model  \citep{2012MNRAS.419.1537B}, but is not constrained either by the short-lived nature of the transient hypermassive \NS\ or by baryon loading of the wind. 
The key point of our model is that a newly formed {\it isolated} \BH\ can retain its magnetic flux,  without any accretion.

\section{Black holes' hair}
\label{hair}

The ``no hair'' theorem  \citep[][]{MTW}, a key result in General Relativity, states that an isolated  black hole  is defined  by  only  three parameters: mass, angular momentum, and  electric charge. We have recently demonstrate that  the ``no hair'' theorem  is not applicable for \BHs\ formed from collapse of a rotating neutron star \citep{2011PhRvD..83l4035L,2011PhRvD..84h4019L}.  
The key point in the classical  proof is that  the outside  medium is a vacuum. In contrast, the surroundings  of astrophysical high energy sources like pulsars and \BHs\ can rarely be treated as vacuum  \citep{GoldreichJulian,Blandford:1977,1992MNRAS.255...61M}.  The ubiquitous presence of  \Bfs\ combined with high (often relativistic) velocities produce inductive \Efs\ with electric potential drops  high enough to break the vacuum via various radiative effects (curvature emission followed by  a single photon pair production in \Bf, or inverse Compton scattering followed by a two photon pair production).  For example, in case of  neutron stars  the rotation of the magnetic field  lines frozen into the crust generates an inductive electric field, which, due to the high conductivity of the neutron star interior,  induces surface  charges. The \Ef\ of these  induced surface   charges has a component parallel to the dipolar \Bf.  These parallel \Efs\ accelerate charges to the energy $ {\cal E} \sim  e B_s R_s ( \Omega R_0/c)^2$, where $B_s$ and $R_s$ are the surface \Bf, radius of a neutron star and $\Omega$  is the angular rotation frequency. The resulting  primary beam of leptons produces a dense secondary plasma via vacuum breakdown. Thus,  in case of neutron stars the electric charges and currents are self-generated: no external source is needed.  Rotating black holes can also lead to a similar vacuum break-down \citep{Blandford:1977}.

Rotating neutron stars can self-produce particles via vacuum breakdown forming a highly conducting  plasma \ms\  with  magnetic field  lines  effectively ``frozen-in'' the star both before and during collapse. 
The electrodynamics of a highly conducting medium is qualitatively different  from the vacuum electrodynamics. The key difference  is that  the highly conducting  plasma quickly shorts out any \Ef\ ($\E$) parallel to magnetic field  ($\B$) through the induction of electric currents \citep[][]{Kulsrud}. The condition $\E\cdot \B=0$ introduces a constraint that  the  magnetic field  lines are effectively frozen into plasma: each plasma element is always ``attached'' to a given magnetic field  line. 
Before the onset of   the collapse,  the electric currents within the neutron star create poloidal \Bf. Rotation of the poloidal magnetic field  lines and the resulting inductive electric field lead to the creation, through vacuum breakdown, of the conducting plasma and poloidal electric currents.  The presence of a conducting plasma then imposes a topological constraint, that the magnetic field  lines which initially were connecting the neutron star surface to the infinity must connect the  black hole  horizon to the infinity.

As a  result, during collapse of a neutron star into a \BH, the latter conserves the number of magnetic flux tubes  $N_B =  e \Phi_\infty /( \pi c \hbar)$, where $\Phi_\infty$ is the initial magnetic flux through the  hemispheres of the progenitor and out to the  infinity. 
 This has been tested  via three-dimensional general relativistic plasma  simulations of rotating black holes that start with a neutron star magnetic field  \citep{2011PhRvD..84h4019L}.

During the collapse, as the surface of a neutron star approaches the horizon,  the closed magnetic field  lines will be quickly absorbed by the \BH, while the open field lines (those connecting to infinity) have to remain open by the frozen-in condition.  Thus, a  black hole  can have only open fields lines, connecting its horizon to the infinity. There is a well known solution that satisfies this condition: an exact  split monopolar solution for rotating \ms\ due to  \cite{Michel73}; it was generalized to Kerr  metrics with small spin parameter   by \cite{Blandford:1977}.  We recently found  an exact non-linear {\it time-dependent }  split monopole-type structure of \mss\  driven by spinning and collapsing neutron star in \Sc\ geometry \citep{2011PhRvD..83l4035L}. We demonstrated that the  collapsing neutron star  enshrouded in a self-generated conducting \ms\ does not allow a  quick release of  the \Bfs\ to infinity \citep{2011PhRvD..84h4019L}. (This result has been recently contested by \cite{2011arXiv1112.2622L}, who claim that \Bfs\ slides off the BH on a dynamical time, and even faster in the presence of plasma; we disagree since the above reference neither  use  conservative MHD numerical schemes nor employ any physical model for the current layer dissipation.)

The discovery  that an isolated  black hole  can retain the magnetic field  opens new possibilities to explain a number of puzzling GRB phenomena that we explore in this paper. The application to GRB is done within the  \EM\ model of short GRBs, which follows the  spirit of our  \EM\ model of GRBs \citep{lb03,Lyutikov:2006b}, that was initially developed for long GRBs. 
The  electromagnetic model assumes that  the rotational energy of a relativistic,  stellar-mass
central source,  is converted into magnetic energy  through unipolar dynamo mechanism,
propagates to large
distances in a form of relativistic
Poynting flux-dominated wind
and is  dissipated directly into emitting particles through current-driven
instabilities. Collimating effects of the magnetic hoop stresses lead to strongly non-spherical
expansion and formation of jets.

\section{Emission phases of short GBRs}
\subsection{The prompt spike}
Merging of \NSs\ is an actively developing area in numerical relativity. At present time we have only   an  overall picture of the merger, but the details, dependences on the parameters of the merger (like the \NSs\ masses), equation of state of the \NSs, effects of the \Bfs\ still remain unclear. In some mergers the two \NSs\ coalesce in the a hyper-massive \NS\ which quickly collapses in the \BH. This scenario is preferred  for massive \NSs\ with soft EoS. For stiffer EoS the merger creates a massive disk \citep[containing up to $0.2 M_\odot$,][]{2010CQGra..27k4105R} that then accretes on the time-scale of hundred milliseconds.

Thus, at the active stage, the 
merger of two \NSs\  first creates a transient hypermassive \NS\  which  within tens to hundreds milliseconds collapses into \BH.
 At the moment of contact the angular velocity and the total kinetic energy of the neutron stars can be estimated as
\ba && 
 \Omega = {1\over 2} \sqrt{ G M_{NS} \over R_{NS}^3} = 6.8 \times 10^3 {\rm rad s^{-1}}
 \label{Omega1}
 \\ &&
 E_{kin} =  M_{NS} v^2 = { G M^2 \over 4 R_{NS}} = 1.3 \times 10^{53} {\rm erg}
 \label{1}
\ea
%\ba &&
%v= {1\over 2} \sqrt{ G M_{NS} \over R_{NS}}
%\nn &&
%\Omega = {1\over 2} \sqrt{ G M_{NS} \over R_{NS}^3} = 6.8 \times 10^3 {\rm rad s^{-1}}
%\nn &&
%I = 2 M v R = \sqrt{ G R_{NS}} M_{NS} ^{3/2}
%\nn &&
%E_{kin} =  M_{NS} v^2 = { G M^2 \over 4 R_{NS}} = 1.3 \times 10^{53} {\rm erg}
%\nn &&
%a= { I \over G M^2/c} = \sqrt{  c^2 R_{NS}  \over G M_{NS}} = 2.19
%\ea
(we assume the conventional $R_{NS} =10 $ km and  $M_{NS}=1.4 \, M_\odot$).  The merger of two NSs creates a  transient hypermassive   NS (HMNS) with $M \sim  2 M_{NS}$ and $R_{HMNS}\sim 2^{1/3} R_{NS}$.
% Once the neutron stars come into contact they have so much angular momentum that they have to form a  an accretion torus with mass $M_d \leq 0.2 M_\odot$. 

\subsection{Generation of \Bf: the HMNS and/or the torus}

The system HMNS-torus may generate \Bf\ in two somewhat related ways. First, the newly born hypermassive NS will be highly sheared, hot and convective and may be surrounded by a heavy $\sim 0.2 M_\odot$ torus \citep{2005PhRvD..71h4021S,2006PhRvL..96c1101D}. These are beneficial conditions for the amplification  of  the \Bf\  via velocity  shear and through dynamo mechanism \citep{1999ApJ...522..753M}. \cite{2006Sci...312..719P} argued that the \Bf\ field can be amplified  {\it within the hypermassive \NS} and sustained at nearly equipartition values of $\sim 10^{16}$ G. 
This exceptionally high level of the \Bf\ amplification within the HMNS  seen by  \cite{2006Sci...312..719P} has been challenged, \eg by  \cite{2011ApJ...732L...6R,2010A&A...515A..30O}, who obtained an alternative scenario \citep[see also][]{2003ApJ...584..937V}): the merger of the \NSs\ results in a formation of the \BH\ (after few tens of millisecond after the \NSs\ come in contact) initially surrounded by 
a hot and highly magnetized torus. The \Bf\ is amplified {\it  within the torus} to  $\sim 10^{15}$ G. The rotating \BH\ is then expected to launch an \EM-jet via the Blandford-Znajek mechanism.

At present, it is not clear which of the two models of the \Bf\ amplification (within the HMNS and/or in the surrounding torus)  gets realized (the answer may also depend on the parameters of the binary, \eg the mass ratio). Below we discuss the general scaling and observations implications for  both possibilities of  the \Bf\ amplification, in the  
BH-torus system (Section \S \ref{BH-torus}) and in the HMNS (Section \S \ref{HMNS}).

\subsubsection{Magnetic field amplification in BH-torus system}
\label{BH-torus}

Let us estimate the \EM\ power produced by the BH-torus system. At the moment of contact the angular momentum of the  \NS\  binary is
$J \sim 2 M_{NS}  \Omega R_{NS}^2$. The Kerr parameter for the resulting BH is then
\be
a= {c\over 4} \sqrt{ R_{NS} \over G M_{NS}} =  \sqrt{R_{NS} \over 4 R_s}  =0.54
\ee
where $R_s = 4 G M_{NS}/c^2 $ is the \Sc\ radius of the resulting BH with mass $M_{BH}= 2 M_{NS}$. 
The resulting BH will have angular velocity
\be
\Omega_H =  { c \over 2 R_+}  = { 2 a \over  1+\sqrt{1-a^2}} {c  \over R_s}
%{ \sqrt{ R_{NS} / R_S} \over 1+ \sqrt{ 1- R_{NS} / (4 R_S)} } \,{ c \over  R_s} 
=2 \times 10^4 {\rm rad s^{-1}}
\label{OmegaH}
\ee
where $R_+ =   (1+\sqrt{1-a^2}) G M /c^2$.

\begin{figure}[h!]
% \vspace{-15pt}
\includegraphics[width=0.99\linewidth]{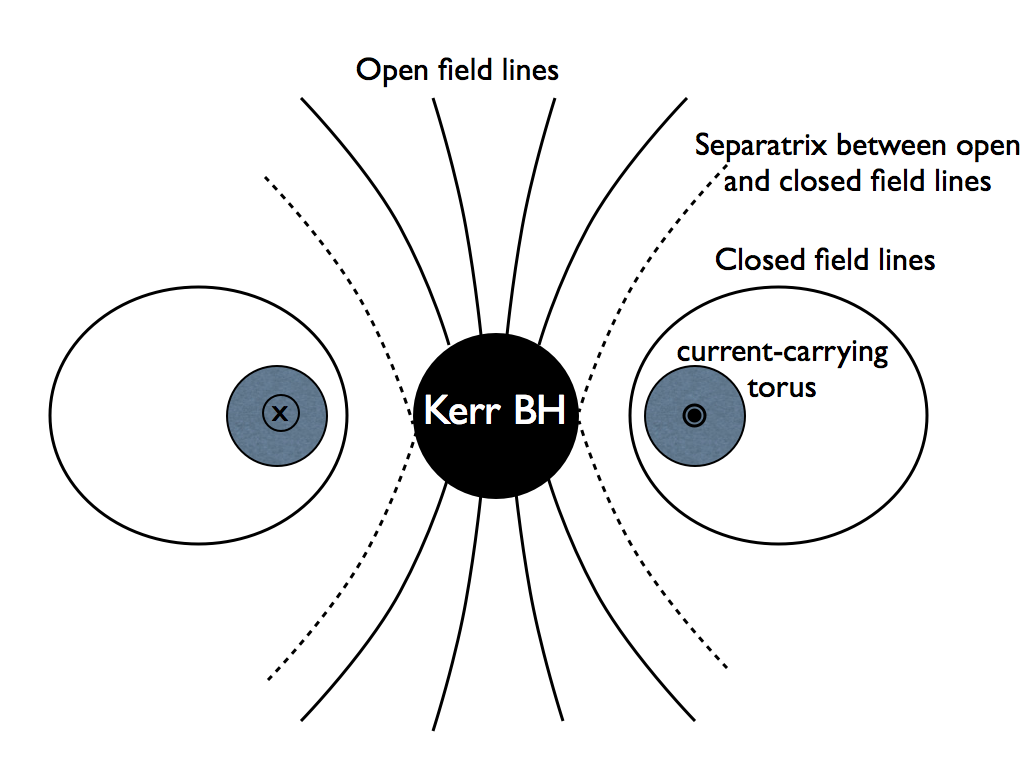}
% \vspace{-35pt}
\caption{Schematic presentations of magnetic flux surfaces in the BH-torus system. Toroidal electric current in the torus creates poloidal \Bf. The field lines that intersect the BH are twisted by the rotation of the space-time (carry poloidal electric current) and open up to infinity.  There are two types of \Bf\ lines separated by a separatrix (dashed lines): closed field lines and open  \Bf\ lines that intersect the Kerr BH. (The section shows only the poloidal component of the \Bf.) After the torus is accreted,  the open \Bf\ lines remain on the BH, relaxing to a twisted  monopolar structure \citep{Michel73,Blandford:1977}. }
\label{torus-pict}
\end{figure}

Assume next that the torus produces \Bf\ $B_T \sim 10^{15} b_{T,15} $ G within its orbit  \citep{2011ApJ...732L...6R}. 
%(For order-of-magnitude estimate of the \Bf, we can assume that the large scale   magnetic field  saturates at  small fraction $\epsilon_B \sim 10^{-5}\epsilon_{B,-5} $ of the kinetic energy, $B^2/(8\pi) (4/3) \pi R^3  = \epsilon_B E_{kin}$ and remains of this order during the merger stage. The small value of the parameter $\epsilon_B$ is expected both due to the loss of energy to gravitational waves and due to the expected low level of magnetic energy at the dynamo  saturation. This gives
%$
%B= { \sqrt{3} \over 2} \sqrt{ \epsilon_B} {\sqrt{G} M_{NS} \over  R_{NS}^2} = 2 \times 10^{15}\, {\rm G} \sqrt{ \epsilon_{B,-5}}  
%$.)
Then the open  magnetic flux through the BH is  \citep[using][ solution]{Wald}, see also \citep{2001bhgh.book.....P}
%Punsly 4.90c
\be 
\Phi_0 = \pi R_+^2 B_T \left( 1 - a^4 \left(    {R_s \over 4  R_+} \right)^4  \right)=
{\pi \over 16} \left(1+\sqrt{1-a^2} \right)^2 \left( 1- \left( {a \over 1+\sqrt{1-a^2}} \right)^4 \right) R_s^2 B_T
\label{Phi01}
\ee
(see Fig. \ref{torus-pict}).

The resulting magnetized \BH\ will produce an 
\EM\ wind with luminosity  \citep{Blandford:1977,2001MNRAS.326L..41K,2005ApJ...630L...5M}
\ba &&
L_{\rm prompt} \approx {2\over 3 c} \left( { \Omega_H\Phi_0 \over 4\pi} \right)^2 = f(a) B_T^2 R_s^2 c= 10^{49} \,  {\rm erg s^{-1}}\, b_{T,15}^2
\nn &&
f(a) = {a^2\over 1536} \left(1+\sqrt{1-a^2} \right)^2 \left( 1- \left( {a \over 1+\sqrt{1-a^2}} \right)^4 \right) 
\label{LNS1}
\ea
For a given mass and \Bf,  the function $f(a)$ reaches maximum at  $a=\sqrt{\frac{\sqrt{5}}{2}-\frac{1}{2}} = 0.78$; for larger Kerr parameters the expulsion of the \Bf\ by the BH reduces the power.

The total wind energy will be determined by the life time of the accretion torus $\tau _T \sim 0.1 {\rm sec} \, \tau _{T,-1}$:
\be
E_{\rm prompt} \approx L_{\rm prompt} \tau _T  =  10^{48} \,  {\rm erg s^{-1}}\,   b_{T,15}^2\, \tau _{T,-1}
\label{Ew}
\ee
This is the typical energy of the short GRBs. Note that it sensitively depends on the assumed level of amplification of the \Bf.

\subsection{Post-collapse evolution: prompt tail}
\label{TTT}

As we discussed in \S \ref{hair} \citep[see also][]{2011PhRvD..83l4035L,2011PhRvD..84h4019L}, the  black hole  resulting from the collapse of a rotating neutron star keeps  for a long time the magnetic field lines that were initially connected to the infinity. Similarly, as the electric current-carrying torus accretes into the BH, the resulting BH will keep the open magnetic flux (\ref{Phi01}).
As a result, the isolated  magnetized \BH\ will produce an 
\EM\ wind with luminosity  (\ref{LNS1}).  But the structure of the wind produced by an isolated \BH\ with monopolar-like \mss\ 
will be different from the jet-like structure produce by the BH-torus system:  isolated  monopolar rotators 
produce {\it equatorially}, not axially collimated wind \citep{Michel73}.

For a moment of inertia of a Kerr  \BH\ $I_{BH}= 2 M^2 R_+$ (in geometric units),  the spin-down time of the \BH\ is fairly long,
\be
\tau_{BH} = {I_{BH} \Omega_H^2 \over L_{BH}}=   {a^2 \over ( 1+\sqrt{1-a^2}) f(a)} {c^3 \over B_T^2 G R_s}= 
10^5\, {\rm sec} \, b_{T,15}^{-2}, 
\label{tauBH}
\ee
 but it sensitively  depends on the assumed level of the \Bf\  amplification. 
 
 Another effect is likely to limit the duration of the post-collapse emission: the resistive loss of the magnetic flux. As \cite{2011PhRvD..84h4019L} discuss, the equatorial current sheet of the monopolar \ms\ is subject
to resistive dissipation that would reconnect the field lines from the
different hemispheres, producing a set of closed field lines that will
be quickly absorbed by the \BH\ and a set of open field lines that
will be released to infinity. This will lead to a decrease of the
number of magnetic flux tubes through each hemisphere, the \BH\
will be slowly ``balding''.  Reconnection of \Bf\ lines is a notoriously difficult problem in
plasma physics \citep[][]{Kulsrud}. Development of plasma turbulence
in the regions of strong current and the resulting anomalous
resistively, plasma collisionality \citep{mu10}, as well as formation
of localized narrow current sheets may bring significant variations
in the dissipation time scale. Using a simplified  Sweet-Parker model of reconnection  \cite{2011PhRvD..84h4019L} estimated that the \Bf\ lines can be retained for hundreds of thousands of the dynamical time scales.   Various reconnection-type phenomena may produce
vastly different time scales which are generally much longer than the dynamical
time scale.

 Thus, the duration and the total energy released by spinning isolated BH depends on the \Bf\ retention time scale, which is likely to be a sensitive function of the microphysics of the relativistic magnetospheric plasma.   
 The total  energy will be determined by  retention time scale $\tau _r \sim 100 {\rm sec} \, \tau _{r,2}$:
\be
E_{\rm tail} \approx L_{\rm tail} \tau _T  =  10^{50} \,  {\rm erg s^{-1}}\,   b_{T,15}^2\, \tau _{r,2}
\label{2} 
  \ee
  Thus, we identify the long prompt emission tails observed in many short GRBs with the emission from the electromagnetic wind generated by the isolated highly spinning \BH.  Note, that the maximal  total energy released in the prompt tail, Eq. (\ref{2}), is larger than the energy of the prompt spike, Eq. (\ref{Ew}).

\section{The \EM\ model of short GRBs}

\subsection{Jetted prompt spike and equatorial prompt tail}

Let us summarize. First, the  two models of the \Bf\ amplification at the prompt stage
   (with \Bf\ amplification with the HMNS and in the surrounding torus) imply different structure of the wind: axially-collimated in case of the BH-torus system 
 \citep[collimated with a typical opening angle of the polar jet of $\sim 30$ degrees][]{2006PhRvL..96c1101D,2011ApJ...732L...6R}
and equatorially--collimated wind if the \Bf\ is generated mostly within the HMNS (and assuming that the corresponding electrodynamics is similar to isolated \NSs). In both cases the resulting isolated BH will produce equatorially--collimated wind. The open magnetic flux is conserved both during the  torus accretion and  during  the collapse 
of the HMNS into the BH; as a result immediately after the collapse the wind power is nearly the same (for the torus-BH system) or even increase for the 
magnetized HMNS (due to BH spin-up during collapse). 

\begin{figure}[h!]
% \vspace{-15pt}
\includegraphics[width=0.99\linewidth]{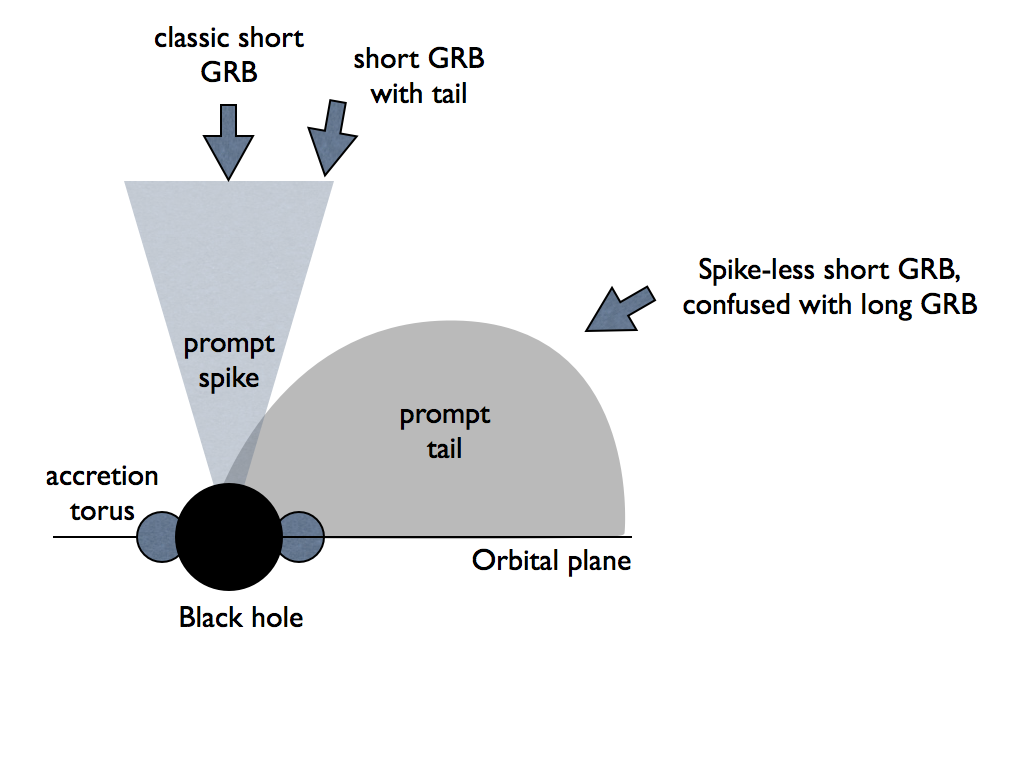}
% \vspace{-35pt}
\caption{Emission diagram for the prompt spike and prompt tails of short GRBs. Merging \NSs\ create a BH-torus system that produces axially aligned jet. Later, after the torus is accreted, an isolated BH produce equatorially collimated wind. If the line of sight  is nearly aligned with the normal to the orbital plane an observer will see a classical short GRB. For medium viewing angles both the prompt spike and the  prompt tail are seen. For large viewing angles the prompt spike is not  seen, so that the burst is mis-identified as a long GRB.  }
\label{GRB-BH-Torus}
\end{figure}

Observational consequences of the merger then highly depend on the viewing angle.
We propose   that it is the BH-torus systems that produce the classical short GRBs  (when viewed nearly along the orbital normal)
and  the short GRBs with prompt tail   (when viewed at some intermediate angle), see Fig. \ref{GRB-BH-Torus}.  If the  line of sight makes a large angle with  the orbital normal we may miss the initial spike, but will still see the tail emission.   Since the 
\Bf\ retention time scale  by the \BH\ is much longer than  torus collapse time, the  tail emission is  intrinsically more energetic. On the other hand, 
the  tail emission is  de-boosted as seen by the observer on the axis, giving a lower observed flux.

 Observers close to the equatorial plane do not see the initial spike, yet  the
 prompt tails have typically been bright enough to be detected with BAT on board Swift  even without the  initial  spike. Such spike-less  prompt tails would be  confused with soft long GRBs. Thus we propose that some bursts identified as   long GBRs are the prompt  tails of the short GRBs, where we missed the initial spike due to misalignment, see Fig. \ref{GRB-BH-Torus}.  Thus, {\it the spike-less short GRBs are confused with some long-soft  GRBs.} (Since 
HMNS collapsing into the BH produce equatorially-collimated flows both at the prompt and tail phases they are mis-identified as long GRBs.)

\subsection{Precursor emission and the Lorimer radio  burst}

Merging \NSs\ are expected to produce a precursor EM wind with power \citep{2001MNRAS.322..695H}
\be
L_p \approx{ ( G M_{NS})^3 B_{NS}^2 \over c^5 R} = { ( G M_{NS})^{9/4} B_{NS}^2 \over c^{15/4}(- t)^{1/4}} = 6\times 10^{43} \, {\rm erg s}^{-1}  \,  {1\over t}  \, b_{12}^2
\label{Lp}
\ee 
where $R$ is the orbital separation, which evolves due to emission of gravitational waves $R= (  (G M)^3 (-t)/c^5)^{1/4}$ for time $-t$ before the merger  \cite[][\S 110]{LLII}.
The stars touch at 
\be t _t\sim -  c^5 R_{NS}^4/( G M_{NS})^3 \approx  - 4\, {\rm msec},
\ee
 at which point the   peak  precursor wind luminosity reaches
\be
L_{p, max}  \approx{ ( G M_{NS})^3 B_{NS}^2 \over c^5 R_{NS}}  = 3 \times 10^{44}  \, {\rm erg s}^{-1}  \, b_{12}^2
\label{LLP}
\ee
This is a fairly low luminosity. Perhaps the only observational possibility to detect the precursor is in the  case when a part of the luminosity (\ref{LLP})  is converted into the pulsar-like coherent radio  emission \citep{2001MNRAS.322..695H} \citep[see also][]{2011MNRAS.415.3065K}. If a fraction $\epsilon_R = 10^{-5} \epsilon_{R,-5}$ of the wind power (\ref{Lp}) is converted into  coherent radio emission, the observed flux at distance $d = 100 d_{100} $ Mpc  and frequency $\nu = 10^9 \, \nu_9$ Hz will be
\be
J= {L \over 4 \pi d^2 \nu} =\epsilon_R  { ( G M_{NS})^3 B_{NS}^2 \over  4 \pi c^5  d^2 \nu (-t/t_t)^{1/4}} = 0.2 \, {\rm Jy}\, (-t/t_t)^{-1/4}\,  \epsilon_{R,-5}\,  \nu_9 \,d_{100} ^{-2}\, b_{12}^2
\label{J}
\ee
These estimates (especially the time scale of few milliseconds) match the so-called Lorimer burst, a highly dispersed impulsive burst-like event with a  duration of few milliseconds \citep{2007Sci...318..777L}. Based on the very large dispersion measure, the burst is cosmological. The rates of  such events, as well as the very validity of the observations remains to be investigated
\citep[][]{2007ApJ...666..346B,2011arXiv1111.0007F,2011ApJ...727...18B} (note that the last observations were done in the continuum at higher frequencies and had much longer timescales, 20 min to 7 days than the Lorimer burst). 
The merging \NS\ model for radio transients has a clear prediction: before the peak the burst luminosity should increase $\propto  (-t/t_t)^{-1/4}$.  A number of such bursts can be detected with LOFAR and the forthcoming  SKA radio telescopes. 

\subsection{Delayed on-set of tail emission}

Typically, the prompt tail emission in short GRBs turn on $\sim $ tens of seconds after the prompt spike. We relate the delayed on-set of tail emission to the effects of Doppler-de-boosting. Let us assume that during the prompt spike a relativistic outflow with total energy $E_0$ has been generated. In addition, after a delay $\Delta t$ another highly relativistic outflow (the prompt tail) is launched. Since the BH-generated winds are expected to be very clean, the velocity of the second outflow is nearly the speed of light. The second outflow will catch with the primary one at time 
\be
t_c \sim \left( { E_0 \Delta t^4 \over c^5 m_p n} \right)^{1/7}
\ee
where $n$ is the circumburst number density. At this moment the Lorentz factor of the  primary Blandford-McKee forward shock is 
\be
\Gamma_c \approx \left( { E_0  \over c^5 m_p n \Delta t^3 } \right)^{1/14} \approx 30 n^{-14} \Delta t^{-3/14} E_{0, 50}^{1/8}
\ee
If viewed at an angle $\theta$, the observed catch-up time is 
\be
t_{c, obs} \approx  \Delta t \left(  1 + \theta^2 \Gamma_c^2\right)
\ee

For example, the BH-torus system is expected to produce a  broad axially  collimated outflows with the opening angle $\theta\sim 10-30$ degrees \citep{2006PhRvL..96c1101D,2011ApJ...732L...6R}. An observer within this angle will see a primary spike. Thus, for a typical axially-aligned observer $\theta \Gamma_c \sim 5-10$, so the tail  on-set time and its duration  $\sim  \Delta t (\theta \Gamma_c )^2 $ is tens of seconds. 
Thus, we relate the tail duration and the delayed on-set due to Doppler de-boosting. This suggestion has a clear 
 predictions for the tails properties: longer tails should statistically have softer spectra and be less luminous.

\section{Nature of flares: episodic magnetized accretion}

Magnetic fields play a crucial role in launching and collimation of magnetized relativistic  outflows. In case of the AGNs and GRBs the \Bfs\ extract the rotational energy of the spinning central  \BH\ \citep{Blandford:1977}. The strength of the \Bf\  near the \BH\ is typically scaled  to a fraction of  the local plasma  rest mass energy density,  $B^2/(8 \pi) \sim \rho c^2$. 

As we demonstrate in this Section, the efficiency of energy extraction of the \BH\ spin energy  during  episodic accretion of magnetized blobs can exceed the average  mass accretion rate $\dot{M} c^2$,  while the total extracted energy can exceed the accreted rest mass. This phenomenon   can lead to production of powerful flares via accretion of fairly small amount of matter.

\subsection{Production of flares}
\begin{figure}[h!]
 \begin{center}
\includegraphics[width=\linewidth]{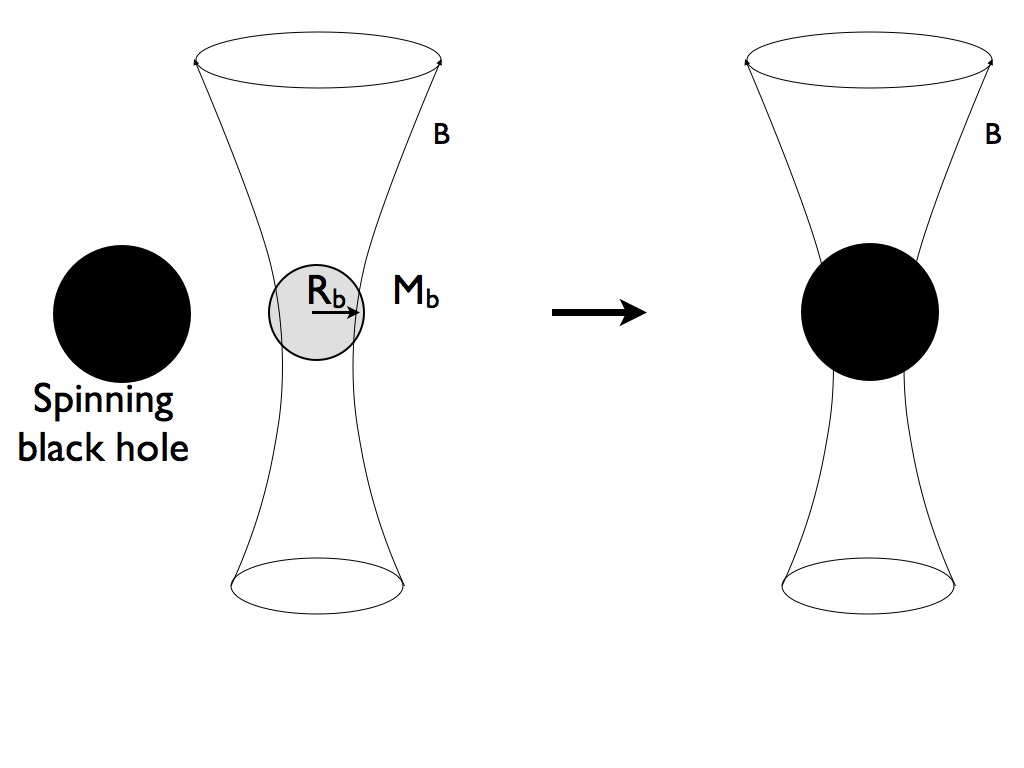}
\end{center}
\caption{Model of episodic accretion. A blob of matter of mass $M_b$ and radius $R_b$ carrying  magnetic flux through it accretes on the spinning  \BH. The \BH\ retains magnetic field for times  much longer than the accretion time, while its rotational energy is extracted electromagnetically through Blandford-Znajek process. For illustrative purposed the overall \Bf\ is assumed to be aligned with the \BH\ spin.}
\label{episodic}
\end{figure}

Let's assume that accretion occurs in a form of magnetized  blobs of mass $M_b$ and radius $R_b$, penetrated by the magnetic flux $\Phi_b = \pi R_b^2 B_b$, see Fig. \ref{episodic}.
These blobs of small total mass were  presumably expelled during the merger of the \NSs. 
The \Bf\ of the blob can be estimated as 
\be
B _b = { 2  \sqrt{2 \pi} \over\beta_b} { G M_b \over R_b^2}
\ee
(this assumes that \Bf\ creates  a pressure scaling as a fraction  $1/\beta_b \leq 1$ of the thermal pressure and that a cloud is thermally supported against its  gravity).
Let us also assume that the  \Bf\  is not confined to the blob  along the field, but extends to infinity, see Fig. \ref{episodic}. The blob carries a magnetic flux $\Phi_b = \pi B_b R_b^2 = 2 \sqrt{2} \pi^{3/2} \sqrt{G} M_b / \sqrt{\beta_b}$.  

After a blob is accreted onto unmagnetized spinning \BH, the \BH\ will posses a  magnetic flux $\Phi_b$, see Fig. \ref{episodic}.  Thus, the \BH\ will produce an \EM\ power (using the BH angular velocity (\ref{OmegaH}))
\be
L_{BH} = {2\over 3 c} \left( { \Omega_H\Phi_b \over 4\pi} \right)^2 = { 4  \pi a^2 \over  3 (1+\sqrt{1-a^2})^2 \beta_b }\,{  c G  M_b^2  \over R_s^2} = 
5\times 10^{47} \,  {1\over \beta_b} \,\left({M_b \over 10^{-5} M_\odot} \right)^{2} \, {\rm erg s}^{-1}
\ee 
(note that the power is independent of the blob's radius $R_b$.)
Thus accretion of a tiny  blob of only $10^{-5} M_\odot$, even accounting for the fact that $\beta_b \geq 1$,  may produce luminosity typical of the GRB flares.

This flux, as we argued above, can be retained for times  $\tau_r$ much longer than the duration of accretion. 
For    long retention time $\tau_r  $ the total  \EM\  energy extracted  electromagnetically from the \BH\   can easily be much larger than the rest mass energy.

We identify the flares observed in GRBs  afterglows with  episodic  accretion events  of magnetized plasma blobs that carry substantial magnetic flux.
The model outlined in this Section  has some similarities to the magnetically switched, rotating black hole model for the FR I/II  galaxies \citep{1999ApJ...522..753M}.

\subsection{Long term \EM\ power}

 Assume next that there are many blobs  that accrete with a rate $\dot{n}$. 
Then the magnetic flux through the hole evolves according to 
\be
\partial_t \Phi =  - \Phi/ \tau_r + \dot{n}\Phi_b 
\ee 
It reaches an  equilibrium value
\be
\Phi_0 =  \dot{n}   \tau_r \Phi_b
\ee

The corresponding Blandford-Znajek  power, 
$
%\be
L_{BZ}= {2\over 3 c} \left( { \Omega_H\Phi_0 \over 4\pi} \right)^2 
%\ee
$
can be much larger (for sufficiently long $\tau_r$)  than the averaged mass accretion rate $\dot{M} = \dot{n}M _b$ 
\be 
{ L_{BZ} \over \dot{M} c^2} =  { 4  \pi a^2 \over  3 (1+\sqrt{1-a^2})^2 \beta  }\,  { \dot{n}   G M_b \tau_r^2 \over c R_s^2}
\ee

\section{Implications}

The \EM\ model of short GRBs  has a number of implications and predictions.  First, it postulates that there are in fact two types of long GRBs, those related to the core-collapse SNe (we refer to them as true long GRBs), and the tails of spike-less short GRBs.  \citep[In fact,][   did speculate that GRB 060614 was a short burst with intense prompt tail]{2006Natur.444.1044G}.  The true  long GRBs should trace the star formation rate (both in terms of the host galaxy types and within a galaxy), while the  spike-less short GRBs should appear in all galaxy types. 
Similarly,  since we relate the SN-less GRBs to short GRBs, they should appear in all galaxy types. The two known SN-less GRBs satisfy this condition: 
the host  of GRB060505 is likely a spiral galaxy, while  the host  of GRB060614 has an order of magnitude  low specific star-formation rate if compared with other nearby long-duration GRB host galaxies  \citep{2006Natur.444.1047F}. 

The different types of  the wind anisotropy at the prompt and the early afterglow stage (axially and equatorially collimated)  bears  implications for afterglow evolution.  We leave a detailed investigation of the afterglow emission to a future investigation. A simple important prediction may be made right away. The equatorial afterglow wind, responsible for the production of the tail, has more energy than the prompt emission.
 At late non-relativistic  stages of the expansion, the axially-aligned observer will infer the explosion energy larger than implied by the prompt emission only. Thus we predict that late afterglows of conventional short GRBs (without tails) should experience re-brightening at the non-relativistic stage. This should be observed in the radio frequency range.

\section{Discussion}

In this paper we outlined the \EM\ model of short GRBs, which follows the  spirit of  our  \EM\ model of  long GRBs \citep{lb03,Lyutikov:2006b}. In particular, we address the key problem of the \NS\ merger paradigm in application to short GRBs, the presence of energetic prompt tails and flares at very long time scales, orders of magnitude longer than the active stage of the merger. 

The key ingredient, that allows the production of the extended outflows from short GRBs is the recent discovery that isolated \BHs\  formed in a collapse of rotating \NSs\ can retain their  open magnetic flux for times much longer than the collapse time and thus can spindown electromagnetically, converting the rotational energy in the \EM\ wind. We identify the prompt GRB spike as coming from the energy  dissipation of the wind powered by a transient accretion torus surrounding the newly formed GRB. It's duration is limited by the life of the torus, tens to hundreds of milliseconds. The long extended emission comes from wind powered by the isolated rotating BH, that produces equatorially-collimated outflow.
It's duration is limited by the retention time scale of the \Bf, and it contains {\it more total energy than the prompt spike}.

Thus, the proposed model for short GRBs implies a different type of collimation of the outflow  than the conventionally envisioned jet-like structure, at least in the prompt tail stage.
 An observer on the axis see only the axially collimated  prompt emission generated by the BH-torus system, while an observer at medium polar angles sees both the prompt spike and the equatorially-collimated extended tail. The flow geometry in principle can be tested using the presumed  ``jet breaks'' - changes in the afterglow light curve when the Lorentz factor of the flow becomes of the order of the opening angle \citep{Rhoads99}. The breaks in case of axially and equatorially collimated outflows should be different. Unfortunately,  the predicted achromatic  light curve breaks  contradict  observation:  often   the breaks are chromatic, with different jumps in optical and X-rays, and sometimes are not observed at all \citep{Panaitescu2007,Racusin09} (see also a discussion by \cite{Lyutikov:2009}). 

We suggest that an observer viewing the merger in the equatorial plane misses the prompt spike, and sees only the extended tail. He then mis-classifies the burst as a long GRB. We propose that  the nearby long burst GRB060505 and GRB060614, that did not show  accompanying supernovae  down to very strict limits \citep{2006Natur.444.1047F}  are, in fact, short GRBs seen in the equatorial plane.

 In addition, we demonstrate that episodic accretion of magnetized, flux-carrying blobs of matter can explain powerful flare observed in GBR afterglows. This effect can also be related to the so-called gigahertz-peaked spectrum  AGN sources, a class of powerful, but short-lived compact radio sources with the intermittent activity of the central engine on time scales as short as thousand years \citep{2009ApJ...698..840C}.

Our \EM\  model of GRBs  explains the surprising similarity between the early afterglows of long and short GRBs: in the \EM\  model both the prompt and the early afterglow are driven by the  fast spinning central black hole, produced through core collapse in long GRBs and  through merger of two \NSs\ in short GRBs. This may also explain the similarity in the GeV signals from short and long GRBs: in the current model they are both produced in the relativistic magnetically dominated \BH\ wind. The very early afterglows in longs GRB can be strongly affected by the accretion of the material from a fall-back disk or from a failed supernova.

The model has a number of the key parameters that are not well constrained: the level of the \Bf\ amplification during a merger, and, most importantly, the retention time scale of the \Bf\ by the \BH. The latter is expected to depend on highly known physics of  relativistic reconnection. In addition, the velocities  of plasma in the \BH\ \ms\ are expected to be strongly sheared, so that the classical reconnection models, that start with a static equilibrium,  will not be applicable in this case. 

I would like to thank Maxim Barkov,  Sarah Burke-Spolaor, Zachary Etienne,  Neil Gehrels, Duncan Lorimer, Andrew McFadyen, Yuk Tun Liu,  Brian Metzger, Rachid Ouyed, Luciano Rezzolla, Stephan Rosswog,  Eduardo  Rubio-Herrera,  Stu Shapiro and organizers of the Aspen conference  The Physics of Astronomical Transients where part of this work was completed. 

 \bibliographystyle{apj}

  \bibliography{/Users/maxim/Home/Research/BibTex}         
 
\appendix

\section{Magnetic field amplification within a hypermassive \NS}
\label{HMNS}

Alternatively, \Bf\ can be amplified within the hypermassive \NS.  
For order-of-magnitude estimates, we can assume that within the HMNS the large scale   magnetic field  saturates at  small fraction $\epsilon_B \sim 10^{-4}\epsilon_{B,-4} $ of the kinetic energy, $B^2/(8\pi) (4/3) \pi R^3  = \epsilon_B E_{kin}$ and remains of this order during the merger stages. The small value of the parameter $\epsilon_B$ is expected both due to the loss of energy to gravitational waves and due to the expected low level of magnetic energy at the dynamo  saturation.
 This gives
\be
B_{\rm HMNS}= { \sqrt{3} \over 2} \sqrt{ \epsilon_B} {\sqrt{G} M_{NS} \over 2 R_{NS}^2} = 6 \times 10^{15}\, {\rm G} \sqrt{ \epsilon_{B,-4}}  
\ee
The structure of the \Bf\ on the HMNS and the corresponding structure of the \EM\ have not been investigated in detail. Let us assume that the \Bf\ both at the inner and outer zones  resembles the dipolar field of conventional pulsar. 
The open magnetic flux through a HMNS is then
\be
\Phi_{\rm HMNS}= \pi R^2 B_{NS} \left({  \Omega R \over c}\right) = { \sqrt{3} \pi \over 2} \sqrt{  \epsilon_B} { G M_{NS}^{3/2} \over c \sqrt{R_{NS}}}
\label{Phi0}
\ee
The transient supermassive neutron star will spin-down electromagnetically, producing  dipolar-like \EM\ wind with luminosity
\be
L_{HMNS} = {4 \over c} \left( { \Omega \Phi_0 \over 4\pi} \right)^2 =  10^{-2}\, \epsilon_B  { G^3 M_{NS}^4 \over c^3 R_{NS}^4} =7
\times 10^{50} {\rm erg s^{-1}}\, \epsilon_{B,-4}
\label{LNS}
\ee

At the HMNS stage the material is expected to be very hot, contaminating the possible \EM\  wind with the neutrino-lifted baryons \citep{2009ApJ...690.1681D}.
The conservation of the magnetic flux  during collapse  may alleviate this problem as the baryons will slide along the \Bf\ lines into the BH, while the BH  keeps its magnetic flux. 
In addition,  as the HMNS rotating with angular velocity (\ref{Omega1}) collapses into the BH rotating with angular velocity (\ref{OmegaH}), it's angular velocity increases:
\be
{\Omega_H \over \Omega} = { 4 \sqrt{2} \over 1+\sqrt{ 1- R_{NS}/(4 R_s)}} \left( {R_{NS} \over R_s} \right)^2 = 4.5
\ee
Since the magnetic flux is nearly conserved on the times scales of the collapse, the emitted power increases by $\sim 20$ during the collapse if compared with the estimate (\ref{LNS}). (Note that this is the total wind power; the  power emitted in the form of high frequency radiation will be a small fraction of the total power.)
The key difference here is that the isolated BH produces equatorially-collimated wind \citep{Michel73}.

\end{document}